\documentstyle[12pt,qqa4lart]{article}
\input tcilatex

\QQQ{Language}{
American English
}

\begin{document}

\author{Lu-Ming Duan and Guang-Can Guo\thanks{%
Electronic address: gcguo@sunlx06.nsc.ustc.edu.cn} \\
Department of Physics and Nonlinear Science Center,\\
University of Science and Technology of China,\\
Hefei 230026, People's Republic of China}
\title{Optimal quantum codes for preventing collective amplitude\\
damping}
\date{}
\maketitle

\begin{abstract}
\baselineskip 24pt Collective decoherence is possible if the departure
between quantum bits is smaller than the effective wave length of the noise
field. Collectivity in the decoherence helps us to devise more efficient
quantum codes. We present a class of optimal quantum codes for preventing
collective amplitude damping to a reservoir at zero temperature. It is shown
that two qubits are enough to protect one bit quantum information, and
approximately $L+\frac 12\log _2\left( \frac{\pi L}2\right) $ qubits are
enough to protect $L$ qubit information when $L$ is large. For preventing
collective amplitude damping, these codes are much more efficient than the
previously-discovered quantum error correcting or avoiding codes.\\

\end{abstract}

\newpage\ \baselineskip 24pt

\section{Introduction}

In quantum computation or communication systems, it is essentially important
to maintain coherence of a quantum system [1]. In reality, however,
decoherence due to the interaction with noisy environment is inevitable [2].
It is discovered that the quantum redundant coding is the most efficient way
to combat decoherence. Till now, many kinds of quantum error correcting or
preventing codes have been devised [3-16]. The quantum error correcting
codes (QECCs) cover a large range of decoherence, and they are very powerful
in noise suppression for large quantum systems. But for small systems, the
QECCs are rather costly of quantum computing resources [17]. To protect one
qubit information from general single-qubit errors, one needs at least five
qubits [7]. Apart from the QECCs, there are alternate quantum codes, such as
the quantum error preventing or avoiding codes [15-20], which combat
decoherence with specific noise modes, but have the advantage of being more
efficient to implement, especially for small quantum systems. The quantum
error preventing codes (QEPCs) are based on the quantum Zeno effect and
therefore useful with quadratic noise [15,16,21]. The quantum error avoiding
codes (QEACs) make use of collectivity in the decoherence [18-20]. For
combatting collective decoherence, they are a better choice.

Collective decoherence is an ideal circumstance, which is possible if the
qubits couple to the same environment, and the separations between them are
smaller than the effective wave length of the noise field. For collective
decoherence, there are coherence preserving states. In the QEACs, arbitrary
input states are encoded into superpositions of the coherence preserving
states. To\ avoid general collective decoherence, one need at least four
qubits to encode one qubit information [20]. Nevertheless, with specific
noise models, more efficient QEACs can be devised. For example, a two-bit
QEAC has been devised for eliminating the dissipation that can be
transformed into collective phase damping by some techniques [19].

The dominant noise process in many quantum computation or communication
systems is described by amplitude damping, such as the radiative decay
[22-25]. In this paper, we propose a class of optimal\ QEACs for preventing
collective amplitude damping to a reservoir at zero temperature. These codes
are much more efficient than those devised in the presence of general
collective decoherence or in the presence of independent amplitude damping
[20,7]. For example, we need only two qubits to encode one qubit
information, and approximately $L+\frac 12\log _2\left( \frac{\pi L}2\right) 
$ qubits to encode $L$ qubit information when $L$ is large. A QEAC with a
high efficiency has two respects of advantages. On the one hand, it costs
few additional quantum computing resources. This is remarkable since quantum
computing resources are very stringent [26,27]. On the other hand, to encode
a bit of information, an efficient QEAC needs only a small number of qubits,
and therefore is much easier to be implemented in practice. The QEACs are
based on collective decoherence. Collective decoherence is most possible for
the closely-spaced adjacent qubits. Cooperative effects in amplitude damping
of two trapped ions have been observed experimentally [28]. In our proposal,
two qubits subject to collective amplitude damping are enough for protecting
one qubit information.

The paper is arranged as follows: First we derive the master equation for
collective amplitude damping. In the derivation, the explicit condition for
collective decoherence is obtained. Then, form the master equation, we show
that there are many collective dark state, which are subjected no collective
amplitude damping. In the whole $2^L$-dimensional Hilbert space of $L$
qubits, the collective dark states span a subspace of dimensions $\left( 
\begin{array}{c}
L \\ 
\left[ L/2\right] 
\end{array}
\right) $, where $\left[ L/2\right] $ indicates the minimum round number no
less than $\frac L2$. For some small $L$, the codes are explicitly
constructed. The $2$-bit code is of special interest, and we further discuss
its possible physical implementation.

\section{The master equation for collective amplitude damping}

We start by deriving the master equation for collective amplitude damping.
Amplitude damping of the qubits is caused by the interaction with noisy
environment. The qubits are described by the spin-$\frac 12$ operators $%
\overrightarrow{s}_l$, and the environment is modeled by a bath of
oscillators with infinite degrees of freedom. The Hamiltonian for amplitude
damping of $L$ qubits in the interaction picture has the following form
(setting $\hbar =1$) 
\begin{equation}
\label{1}H_I\left( t\right) =\stackrel{L}{\stackunder{l=1}{\sum }}%
\stackunder{\overrightarrow{k}}{\sum }\left[ g_{\overrightarrow{k}}e^{-i%
\overrightarrow{k}\cdot \overrightarrow{r}_l}e^{-i\left( \omega _{%
\overrightarrow{k}}-\omega _0\right) t}s_l^{+}a_{\overrightarrow{k}%
}+H.c.\right] 
\end{equation}
where $a_{\overrightarrow{k}}$ is the annihilation operator of the bath mode 
$\overrightarrow{k}$, and $\omega _{\overrightarrow{k}}$ and $\omega _0$
denote frequencies of the bath mode $\overrightarrow{k}$ and of the qubits,
respectively. The symbol $\overrightarrow{r}_l$ indicates the site of the $l$
qubit, and $g_{\overrightarrow{k}}$ is the coupling coefficient. Under the
Born-Markov approximation, the general form of the master equation with the
interaction Hamiltonian $H_I\left( t\right) $ is expressed as [29] 
\begin{equation}
\label{2}\frac d{dt}\rho \left( t\right) =-\int_0^\infty d\tau \text{tr}%
_B\left\{ \left[ H_I\left( t\right) ,\left[ H_I\left( t-\tau \right) ,\rho
\left( t\right) \otimes \rho _B\right] \right] \right\} , 
\end{equation}
where $\rho _B$ is the bath density operator, and $\rho \left( t\right) $
denotes the reduced density operator of the qubits in the interaction
picture. Suppose that the bath is at zero temperature. This is the case in
many circumstances, such as for the radiative decay or for the loss process
[22-25]. Substituting the Hamiltonian (1) into Eq. (2) , we get the
following master equation for spatially-correlated amplitude damping 
\begin{equation}
\label{3}\frac d{dt}\rho \left( t\right) =i\stackrel{L}{\stackunder{i,j=1}{%
\sum }}\delta _{ij}\left[ s_j^{+}s_i^{-},\rho \left( t\right) \right] +\frac
12\stackrel{L}{\stackunder{i,j=1}{\sum }}\left\{ \gamma _{ij}\left[
2s_i^{-}\rho \left( t\right) s_j^{+}-s_j^{+}s_i^{-}\rho \left( t\right)
-\rho \left( t\right) s_j^{+}s_i^{-}\right] \right\} , 
\end{equation}
where the spatially-correlated damping coefficients $\gamma _{ij}$ and Lamb
shifts $\delta _{ij}$ are defined respectively by 
\begin{equation}
\label{4}\gamma _{ij}=\stackunder{\overrightarrow{k}}{\sum }\left[ 2\pi
\left| g_{\overrightarrow{k}}\right| ^2\delta \left( \omega _{%
\overrightarrow{k}}-\omega _0\right) e^{i\overrightarrow{k}\cdot \left( 
\overrightarrow{r}_i-\overrightarrow{r}_j\right) }\right] , 
\end{equation}

\begin{equation}
\label{5}\delta _{ij}=\stackunder{\overrightarrow{k}}{\sum }\left[ \left| g_{%
\overrightarrow{k}}\right| ^2\frac 1{\omega _{\overrightarrow{k}}-\omega
_0}e^{i\overrightarrow{k}\cdot \left( \overrightarrow{r}_i-\overrightarrow{r}%
_j\right) }\right] .
\end{equation}
In the continuum limit, the summations of Eqs. (4) and (5) become integrals
and the principal should be taken of the integral of Eq. (5). The main
contributions to the summations of Eqs. (4) and (5) come form the modes $%
\overrightarrow{k}$ that satisfy $\omega _{\overrightarrow{k}}\approx \omega
_0$. Suppose $d$ is the maximum separation between the qubits, and $v_0$ is
the velocity of the noise field around $\omega _{\overrightarrow{k}}=\omega
_0$, i.e., $v_0=\left. \frac{\omega _{\overrightarrow{k}}}{\left| 
\overrightarrow{k}\right| }\right| _{\omega _{\overrightarrow{k}}=\omega _0}$%
. If $d$ and $v_0$ satisfy the condition 
\begin{equation}
\label{6}d<<\frac{v_0}{\omega _0},
\end{equation}
in Eqs. (4) and (5) $e^{i\overrightarrow{k}\cdot \left( \overrightarrow{r}_i-%
\overrightarrow{r}_j\right) }\approx 1$, and then $\gamma _{ij}$ and $\delta
_{ij}$ are independent of the qubit index. In this circumstance, we denote $%
\gamma _{ij}=\gamma _0$, $\delta _{ij}=\delta _0$, and $S^{\pm }=\stackrel{L%
}{\stackunder{l=1}{\sum }}s_l^{\pm }$. Eq. (3) is thus simplified to 
\begin{equation}
\label{7}\frac d{dt}\rho \left( t\right) =i\delta _0\left[ S^{+}S^{-},\rho
\left( t\right) \right] +\frac{\gamma _0}2\left[ 2S^{-}\rho \left( t\right)
S^{+}-S^{+}S^{-}\rho \left( t\right) -\rho \left( t\right) S^{+}S^{-}\right]
.
\end{equation}
This is the master equation for collective amplitude damping, which is
obtained under the condition (6). The term $\frac{v_0}{\omega _0}$ in Eq.
(6) defines the effective wave length of the noise field. This expression
for the effective wave length is gained under the Born-Markov approximation,
and holds in the case of amplitude damping. For other sources of
decoherence, the expression for the effective wave length may have a
different form [30]. The condition (6) may be satisfied in practice for some
sources of decoherence. For example, in the ion trap quantum computer, a
fundamental limit to internal state decoherence is given by the radiative
decay. For this source of decoherence, $v_0$ is estimated by the velocity of
light, and the typical value of the separations of ions (qubits) has the
order of a few $\mu m$, then Eq. (6) requires that $\omega _0<<10^{14}Hz$.
For some hyperfine transitions, it is possible to meet this condition [27].

\section{Collective dark states}

In the language of quantum trajectories [31], the system evolution described
by the master equation (7) is represented by an ensemble of wave functions
that propagate according to the effective Hamiltonian 
\begin{equation}
\label{8}H_{eff}=-\delta _0S^{+}S^{-}-\frac i2\gamma _0S^{+}S^{-}, 
\end{equation}
interrupted at random times by quantum jumps. A\ quantum jump takes place in
the time interval $[t,t+dt)$ with probability 
\begin{equation}
\label{9}P\left( t\right) =\left\langle \Psi \left( t\right) \right| \gamma
_0S^{+}S^{-}\left| \Psi \left( t\right) \right\rangle dt, 
\end{equation}
leading to a wave function collapse according to 
\begin{equation}
\label{10}\left| \Psi \left( t+dt\right) \right\rangle =c^{^{\prime }}\sqrt{%
\gamma _0}S^{-}\left| \Psi \left( t\right) \right\rangle , 
\end{equation}
where $c^{^{\prime }}$ is a normalization constant. From Eqs. (8) and (9) it
follows that if a initial state satisfies 
\begin{equation}
\label{11}S^{-}\left| \Psi \left( 0\right) \right\rangle =0, 
\end{equation}
it remains unchanged during the effective evolution, and is subjected to no
quantum jumps at any time. All the states satisfying Eq. (11) are called the
collective dark states. Coherence between these states is perfectly
preserved during collective amplitude damping. It can also be seen from Eqs.
(8) and (9) that no other states except those satisfying Eq. (11) remain
unchanged during the effective evolution and quantum jumps.

To get all the collective dark states, we notice that $\overrightarrow{S}=%
\stackrel{L}{\stackunder{l=1}{\sum }}\overrightarrow{s}_l$ is expressed as a
sum of $L$ spin-$\frac 12$ operators. From the angular momentum theory [32], 
$S^{\left( x\right) },S^{\left( y\right) },$ and $S^{\left( z\right) }$ can
be chosen as three generators of the $su(2)$ algebra. The irreducible
representation of the $su(2)$ algebra in the $2$-dimensional Hilbert space $%
H_{\frac 12}$ of a single qubit is denoted by $D_{\frac 12}$, then $D_{\frac
12}^{\otimes L}$ defines an $L$-fold tensor product representation of the $%
su(2)$ algebra in the whole $2^L$-dimensional Hilbert space $H_{\frac
12}^{\otimes L}$ of $L$ qubits. The representation $D_{\frac 12}^{\otimes L}$
is reducible, and it can be decomposed into a series of irreducible
representations of the $su(2)$ algebra, such as 
\begin{equation}
\label{12}D_{\frac 12}^{\otimes 2}=D_{\frac 12}\otimes D_{\frac
12}=D_1\oplus D_0.
\end{equation}
Suppose $D_{\frac 12}^{\otimes 2l}$ has the decomposition $D_{\frac
12}^{\otimes 2l}=\stackunder{j=0}{\stackrel{l}{\oplus }}n_j\left( 2l\right)
D_j\left( 2l\right) $, where $D_j\left( 2l\right) $ denotes the $(2j+1)$%
-dimensional irreducible representations of the $su(2)$ algebra in the state
space of $2l$ qubits, and $n_j\left( 2l\right) $ is the multiplicity of $%
D_j\left( 2l\right) $ in the decomposition, then we have the following
recursion relations (setting $n_{-1}\left( 2l\right) =n_{l+1}\left(
2l\right) =n_{l+2}\left( 2l\right) =0$) 
\begin{equation}
\label{13}D_{\frac 12}^{\otimes 2l+1}=D_{\frac 12}^{\otimes 2l}\otimes
D_{\frac 12}=\stackunder{j=\frac 12}{\stackrel{l+\frac 12}{\oplus }}\left[
n_{j+\frac 12}\left( 2l\right) +n_{j-\frac 12}\left( 2l\right) \right]
D_j\left( 2l+1\right) ,
\end{equation}

\begin{equation}
\label{14}D_{\frac 12}^{\otimes 2l+2}=D_{\frac 12}^{\otimes 2l+1}\otimes
D_{\frac 12}=\stackunder{j=0}{\stackrel{l+1}{\oplus }}\left[ 2n_j\left(
2l\right) +n_{j-1}\left( 2l\right) +n_{j+1}\left( 2l\right) \right]
D_j\left( 2l+2\right) . 
\end{equation}
Equations (13) and (14), together with Eq. (12), determine the decomposition
of $D_{\frac 12}^{\otimes L}$ with an arbitrary $L$. In the decomposition of 
$D_{\frac 12}^{\otimes L}$, there are $n_j\left( L\right) $ $(2j+1)$%
-dimensional irreducible representations $D_j\left( L\right) $, whose
representation spaces are denoted by $H_j^{\left( m\right) }\left( L\right) $%
, where $m=1,$ $2,$ $\cdots ,$ and $n_j\left( L\right) $, respectively. The
whole $2^L$-dimensional Hilbert space $H_{\frac 12}^{\otimes L}$ of $L$
qubits splits into a series of orthogonal subspaces $H_j^{\left( m\right)
}\left( L\right) $ according to the decomposition of $D_{\frac 12}^{\otimes
L}$. In every subspace $H_j^{\left( m\right) }\left( L\right) $, the Casimir
operator $\overrightarrow{S}^2=\left( S^{\left( x\right) }\right) ^2+\left(
S^{\left( y\right) }\right) ^2+\left( S^{\left( z\right) }\right) ^2$ has
the eigenvalue $j(j+1)$. The subspace $H_j^{\left( m\right) }\left( L\right) 
$ is of $2j+1$ dimensions, whose basis-vectors can be chosen as the
eigenvectors $\left| j,m_j\right\rangle _m$ of the operator $S^{\left(
z\right) }$, where $m_j=-j,-j+1,\cdots ,j$. In each space $H_j^{\left(
m\right) }\left( L\right) $, the lowest-weight state $\left|
j,-j\right\rangle _m$ satisfies the condition $S^{\left( -\right) }\left|
j,-j\right\rangle _m=0$, and no other states have this property. Hence there
is one and merely one collective dark state in each subspace $H_j^{\left(
m\right) }\left( L\right) $, and the dark states in different subspaces are
orthogonal to each other. The total number $N\left( L\right) $ of orthogonal
collective dark states is therefore just the number of the irreducible
representations in the decomposition of $D_{\frac 12}^{\otimes L}$, i.e.,
the total number $N\left( L\right) =\sum_jn_j\left( L\right) $. From Eqs.
(13) and (14), we get the following recursion equations about $N\left(
L\right) $%
\begin{equation}
\label{15}N\left( 2l+1\right) =2N\left( 2l\right) -n_0\left( 2l\right) , 
\end{equation}
\begin{equation}
\label{16}N\left( 2l+2\right) =2N\left( 2l+1\right) , 
\end{equation}
where $n_0\left( 2l\right) $ is the multiplicity of the $1$-dimensional
irreducible representations in the decomposition of $D_{\frac 12}^{\otimes
L} $, and is known to be $n_0\left( 2l\right) =\left( 2l\right) !\left[
l!\left( l+1\right) !\right] ^{-1}$ [20]. Substituting it into Eqs. (15) and
(16), we get $N\left( L\right) =\left( 
\begin{array}{c}
L \\ 
\left[ L/2\right] 
\end{array}
\right) $, where $\left[ L/2\right] $ indicates the minimum round number no
less than $\frac L2$. The quantum error avoiding codes are obtained by
encoding arbitrary input states into superpositions of the collective dark
states. The encoding space is of $N\left( L\right) $ dimensions, thus the
optimal $L$-bit quantum code has the efficiency 
\begin{equation}
\label{17}\eta \left( L\right) =\frac 1L\log _2N\left( L\right) =\frac
1L\log _2\left( 
\begin{array}{c}
L \\ 
\left[ L/2\right] 
\end{array}
\right) . 
\end{equation}
If $L$ is large, $\eta \left( L\right) $ is approximated by $1-\frac
1{2L}\log _2\left( \frac{\pi L}2\right) $, which approaches $1$ very
rapidly. Hence, in the presence of collective amplitude damping, these codes
are much more efficient than the previously-discovered quantum error
correcting or avoiding codes.

\section{Explicit constructions of the $L$-bit codes with some small $L$}

The orthogonal collective dark states obtained in the previous section can
be chosen as a set of basis-vectors for the encoding space. To explicitly
construct the codes, we need only express the collective dark states in the
computation basis, whose basis-vectors are the co-eigenstates of the
operators $s_1^z$, $s_2^z,$ $\cdots ,$ and $s_L^z$. The two eigenstates of
the operator $s_l^z$, with the eigenvalues $\pm \frac 12$, are denoted by $%
\left| 1\right\rangle $ and $\left| 0\right\rangle $, respectively. The
collective dark states and the computational basis-vectors are connected by
the Clebsch-Gordan coefficients [32]. Here, we explicit construct the
optimal $L$-bit QEACs with $L=2,3,4$. These codes are simple and involve
only a small number of qubits, and at the same time have notably high
efficiencies, so they are an ideal choice of quantum codes in the presence
of collective amplitude damping.

In the case of two qubits, the encoding space is of two dimensions. The two
codewords are given by 
\begin{equation}
\label{18}\left| j=0,m_j=0\right\rangle =\frac 1{\sqrt{2}}\left( \left|
01\right\rangle -\left| 10\right\rangle \right) , 
\end{equation}
\begin{equation}
\label{19}\left| j=1,m_j=-1\right\rangle =\left| 00\right\rangle , 
\end{equation}
which are sufficient to encode one qubit information. The efficiency is $%
\frac 12$.

In the case of three qubits, the encoding space is of three dimensions. The
codewords read 
\begin{equation}
\label{20}\left| j=\frac 12,m_j=-\frac 12\right\rangle _1=\frac 1{\sqrt{6}%
}\left( \left| 001\right\rangle +\left| 100\right\rangle -2\left|
010\right\rangle \right) , 
\end{equation}
\begin{equation}
\label{21}\left| j=\frac 12,m_j=-\frac 12\right\rangle _2=\frac 1{\sqrt{2}%
}\left( \left| 001\right\rangle -\left| 100\right\rangle \right) , 
\end{equation}
\begin{equation}
\label{22}\left| j=\frac 32,m_j=-\frac 32\right\rangle =\left|
000\right\rangle . 
\end{equation}
The efficiency of this code is $\frac 13\log _23$. At least one qubit
information can be encoded.

In the case of four qubits, the encoding space is of six dimensions. The
codewords are respectively 
\begin{equation}
\label{23}\left| j=0,m_j=0\right\rangle _1=\frac 12\left( \left|
01\right\rangle -\left| 10\right\rangle \right) \left( \left|
01\right\rangle -\left| 10\right\rangle \right) , 
\end{equation}
\begin{equation}
\label{24}\left| j=0,m_j=0\right\rangle _2=\frac 1{\sqrt{3}}\left[ \left|
0011\right\rangle +\left| 1100\right\rangle -\frac 12\left( \left|
01\right\rangle +\left| 10\right\rangle \right) \left( \left|
01\right\rangle +\left| 10\right\rangle \right) \right] , 
\end{equation}
\begin{equation}
\label{25}\left| j=1,m_j=-1\right\rangle _1=\frac 1{\sqrt{2}}\left( \left|
01\right\rangle -\left| 10\right\rangle \right) \left| 00\right\rangle , 
\end{equation}
\begin{equation}
\label{26}\left| j=1,m_j=-1\right\rangle _2=\frac 1{\sqrt{2}}\left|
00\right\rangle \left( \left| 01\right\rangle -\left| 10\right\rangle
\right) , 
\end{equation}
\begin{equation}
\label{27}\left| j=1,m_j=-1\right\rangle _3=\frac 12\left[ \left( \left|
01\right\rangle +\left| 10\right\rangle \right) \left| 00\right\rangle
-\left| 00\right\rangle \left( \left| 01\right\rangle +\left|
10\right\rangle \right) \right] , 
\end{equation}
\begin{equation}
\label{28}\left| j=2,m_j=-2\right\rangle =\left| 0000\right\rangle . 
\end{equation}
The efficiency of this code is $\frac 14\left( 1+\log _23\right) $. At least
two qubit information can be encoded.

The $2$-bit code is of special interest. It costs least number of qubits,
and therefore has a good chance to be first implemented. We further give the
encoding and decoding for this code. Let $C_{ij}$ and $C_{ij}\left( H\right) 
$ denote the controlled-Not and the controlled-Hadamard operations,
respectively, where the first subscript of $C_{ij}$ or $C_{ij}\left(
H\right) $ refers to the control bit and the second to the target. The
controlled Hadamard operation performs the Hadamard transformation ($\left|
1\right\rangle \rightarrow \left( \left| 1\right\rangle +\left|
0\right\rangle \right) /\sqrt{2},$ $\left| 0\right\rangle \rightarrow \left(
\left| 1\right\rangle -\left| 0\right\rangle \right) /\sqrt{2}$) on the
target bit if the control bit is in$\left| 1\right\rangle $, and leaves the
target bit unchanged if the control bit is in $\left| 0\right\rangle $. The
input state of a single qubit can be generally expressed as $\left| \Psi
\left( 0\right) \right\rangle _1=c_0\left| 0\right\rangle +c_1\left|
1\right\rangle $. An ancillary qubit $2$ is pre-arranged in the state $%
\left| 0\right\rangle _2$. The input state is encoded by the following
operation 
\begin{equation}
\label{29}\left| \Psi \left( 0\right) \right\rangle _1\left| 0\right\rangle
_2\stackrel{C_{21}C_{12}\left( H\right) }{\longrightarrow }\left| \Psi
_{enc}\right\rangle _{12}=c_0\left| 00\right\rangle +\frac{c_1}{\sqrt{2}}%
\left( \left| 01\right\rangle -\left| 10\right\rangle \right) . 
\end{equation}
The encoded state is subjected to no collective amplitude damping, and
afterwards it can be decoded by applying the same operation again in the
reverse order, i.e., 
\begin{equation}
\label{30}\left| \Psi _{enc}\right\rangle _{12}\stackrel{C_{12}\left(
H\right) C_{21}}{\longrightarrow }\left| \Psi \left( 0\right) \right\rangle
_1\left| 0\right\rangle _2. 
\end{equation}
The controlled-NOT and the controlled-Hadamard operations involved in the
encoding and decoding have been demonstrated [26,27], and cooperative
effects in amplitude damping of two trapped ions have been observed
experimentally [28], so the proposed $2$-bit code has a good chance to be
implemented in the near future experiment.\\

{\bf Acknowledgment}

This project was supported by the National Nature Science Foundation of
China.

\newpage\

\end{document}